\documentclass[sigconf,natbib=true]{acmart}
\usepackage{multirow} 
\usepackage{enumitem}
\usepackage{makecell}
\usepackage{algorithm}
\usepackage{algorithmic}

\AtBeginDocument{%
  }

\copyrightyear{2025}
\acmYear{2025}
\setcopyright{acmlicensed}\acmConference[SIGIR '25]{Proceedings of the 48th
International ACM SIGIR Conference on Research and Development in
Information Retrieval}{July 13--18, 2025}{Padua, Italy}
\acmBooktitle{Proceedings of the 48th International ACM SIGIR Conference on
Research and Development in Information Retrieval (SIGIR '25), July 13--18,
2025, Padua, Italy}
\acmDOI{10.1145/3726302.3729932}
\acmISBN{979-8-4007-1592-1/2025/07}

\begin{document}

\title{Comprehending Knowledge Graphs with Large Language Models for Recommender Systems}

\author{Ziqiang Cui}
\authornote{Work done during an internship at Tencent.}
\affiliation{
  \institution{City University of Hong Kong}
  \country{Hong Kong SAR}
}
\email{ziqiang.cui@my.cityu.edu.hk}

\author{Yunpeng Weng}
\affiliation{
  \institution{FiT, Tencent}
  \city{Shenzhen}
  \country{China}
}
\email{edwinweng@tencent.com}

\author{Xing Tang}
\authornotemark[2]
\affiliation{
  \institution{FiT, Tencent}
  \city{Shenzhen}
  \country{China}
}
\email{xing.tang@hotmail.com}

\author{Fuyuan Lyu}
\affiliation{
  \institution{McGill University \& MILA}
  \city{Montréal}
  \country{Canada}
}
\email{fuyuan.lyu@mail.mcgill.ca}

\author{Dugang Liu}
\affiliation{
  \institution{Shenzhen University}
  \city{Shenzhen}
  \country{China}
}
\email{dugang.ldg@gmail.com}

\author{Xiuqiang He}
\affiliation{
  \institution{Shenzhen Technology University}
  \city{Shenzhen}
  \country{China}
}
\email{hexiuqiang@qq.com}

\author{Chen Ma}
\authornote{Corresponding author.}
\affiliation{
  \institution{City University of Hong Kong}
  \country{Hong Kong SAR}
}
\email{chenma@cityu.edu.hk}

\begin{abstract} 
In recent years, the introduction of knowledge graphs (KGs) has significantly advanced recommender systems by facilitating the discovery of potential associations between items. However, existing methods still face several limitations. First, most KGs suffer from missing facts or limited scopes. Second, existing methods convert textual information in KGs into IDs, resulting in the loss of natural semantic connections between different items. Third, existing methods struggle to capture high-order connections in the global KG.
To address these limitations, we propose a novel method called \textbf{CoLaKG}, which leverages large language models (LLMs) to improve KG-based recommendations. The extensive knowledge and remarkable reasoning capabilities of LLMs enable our method to supplement missing facts in KGs, and their powerful text understanding abilities allow for better utilization of semantic information.
Specifically, CoLaKG extracts useful information from KGs at both local and global levels. By employing the item-centered subgraph extraction and prompt engineering, it can accurately understand the local information. In addition, through the semantic-based retrieval module, each item is enriched by related items from the entire knowledge graph, effectively harnessing global information. Furthermore, the local and global information are effectively integrated into the recommendation model through a representation fusion module and a retrieval-augmented representation learning module, respectively. Extensive experiments on four real-world datasets demonstrate the superiority of our method.
The code of our method is available at \textcolor{blue}{\url{https://github.com/ziqiangcui/CoLaKG}}.
\end{abstract}



\maketitle

\section{Introduction}

The rapid advancement of online platforms has led to an increasingly critical issue of information overload.  Recommender systems address this problem by modeling user preferences based on historical data. Collaborative filtering (CF) \cite{rendle2012bpr,he2017neural,he2020lightgcn}, as one of the most classic and efficient methods, has been extensively employed in existing recommender systems. However, CF-based methods exclusively rely on user-item interaction records, often suffering from the data sparsity issue \cite{togashi2021alleviating}. 
To address this issue, recent studies \cite{wang2019kgat,zhang2016collaborative,yang2022knowledge} have incorporated knowledge graphs (KGs) as external knowledge sources into recommendation models, achieving significant progress. Typically, these methods capture diverse and high-order relationships between items by modeling the structure and attribute information in KGs, thereby enhancing the learning process of user and item representations \cite{yang2022knowledge}.

\begin{figure}[t]
\setlength{\belowcaptionskip}{-5mm} 
  \centering  
  \includegraphics[width=0.85\columnwidth]{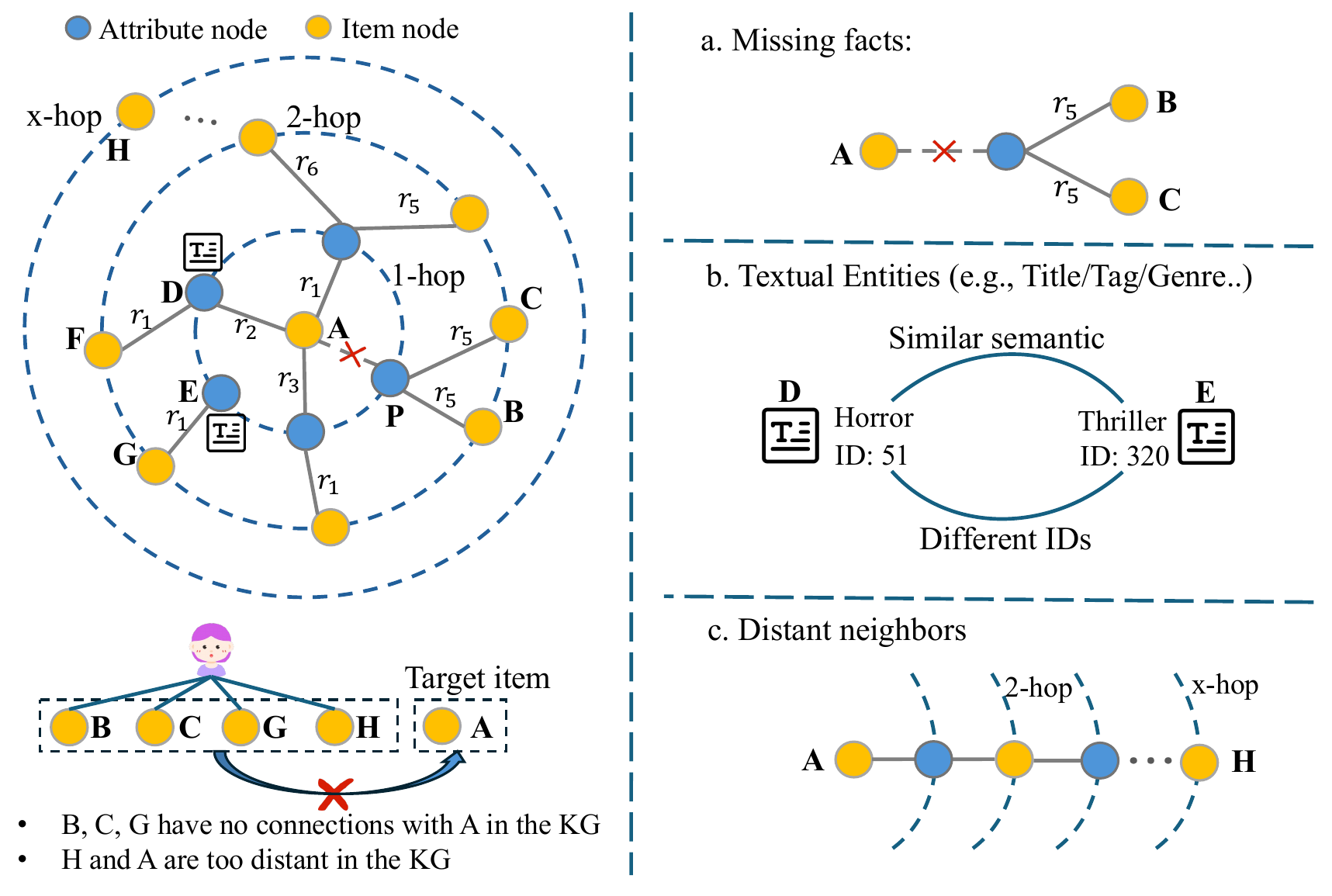}
  \caption{An illustrative diagram demonstrating the potential issues of existing KG-based recommendation methods.} 
   \label{intro_fig}
   \vspace{-0.5em}
\end{figure}
Despite the effectiveness of existing KG-enhanced recommendation methods, they still face several challenges.
i) First, many KGs suffer from missing facts and limited scopes \cite{guo2020survey,wang2023federated}, as constructing KGs often requires manual effort and domain expertise. The absence of key attributes, such as the genres of a movie, can cause items that originally share the same attribute to lose their connections. As illustrated in Figure \ref{intro_fig}, items A and B should have a connecting path (A-P-B) in the KG. However, due to item A missing the P attribute, A and B are now not associated with each other. Such biased knowledge will lead to suboptimal performance of the recommendation model.
ii) Second, existing methods \cite{yang2022knowledge,yang2023knowledge,wang2019kgat,wang2021learning,wang2024unleashing} convert textual entities and relations into IDs for use, which can result in losing natural semantic connections between items. For instance, in Figure \ref{intro_fig}, ``horror'' and ``thriller'' are two semantically related attributes of Item F and Item G, respectively. 
However, similar semantics are not reflected in different entity IDs (51 and 320), which further results in the related items F and G having no connections in the KG.
iii) Third, existing methods \cite{wang2019kgat,yang2022knowledge,yang2023knowledge,wang2021learning,wang2024unleashing,tian2021joint} struggle to capture high-order relationships in the entire KG. Most of them propagate and aggregate information by stacking multiple layers of graph neural networks (GNNs). The layer-by-layer propagation is not only inefficient but also accumulates a large amount of irrelevant node information, leading to the over-smoothing issue \cite{guo2020survey,wang2024unleashing}. 
For instance, assumming that points A and H in Figure \ref{intro_fig} have a strong semantic connection, the considerable distance between them in the KG presents big challenges for existing methods in capturing this relation. 
As shown in the lower left corner of Figure \ref{intro_fig}, due to the aforementioned three limitations, existing methods fail to identify the potential connections between historically interacted items and the target item through the KG, resulting in suboptimal recommendations.

Empowered by extensive knowledge and remarkable reasoning abilities, large language models (LLMs) have demonstrated significant promise in semantic understanding and knowledge extraction. Recently, many studies have leveraged LLMs to improve recommendation models. For example, some studies utilize LLMs to generate semantic representations of item profiles \cite{ren2024representation}, while other studies \cite{zhao2024breaking} employ LLMs to determine whether a complementary relationship exists between two items, thereby recommending complementary products based on users' historical behaviors.
However, these methods do not fully exploit the semantic and structural information of KGs. As one of the most common and important sources of knowledge, KGs contain a wealth of semantic associations among entities and relations, which are often overlooked by existing methods that typically consider only item profiles. Additionally, KGs serve as task-relevant knowledge repositories, effectively aiding LLMs in acquiring task-specific knowledge and mitigating the issue of hallucinations caused by excessive divergence.
Nevertheless, leveraging LLMs to handle the diverse semantic relationships in KGs is highly challenging, as it is unrealistic to input the vast number of entities and connections in KGs into a language model.

To bridge this gap, in this paper, we propose a novel method named \textbf{Co}mprehending \textbf{K}nowledge \textbf{G}raphs with \textbf{La}rge Language Models for Recommendation (\textbf{CoLaKG}). The core idea is to leverage LLMs for understanding the semantic and structural information of KGs to enhance the representation learning of items and users. 
Our method comprises two stages: 
i) \textit{Comprehending KGs with LLMs.} Given the impracticality of inputting the entire KG into an LLM, CoLaKG leverages KG information by two components: a \textbf{Local KG Comprehension} module which involves extracting item-centered KG subgraphs for analysis by LLMs, and a \textbf{Retrieval-based Global KG Utilization} module which retrieves semantically related items from the entire KG to effectively leverage global KG information.
ii) \textit{Incorporating semantic embeddings into the recommendation model.} 
This stage integrates the semantic embeddings of KGs with the ID embeddings, thereby leveraging both collaborative signals and semantic information. First, we align and fuse the item ID representations with the item-centered KG subgraph representations. Then, we employ a retrieval-augmented representation method to further enhance item representations with semantically similar neighbors from the entire KG.

It is important to note that these two stages are decoupled, meaning that our method does not involve LLM inference during the recommendation process. This allows our method to be efficiently applied in real-world recommendation scenarios.

Our contributions are summarized as follows:
\begin{itemize}[leftmargin=1em,topsep=2pt,parsep=2pt]
\item We propose a novel method that utilizes LLMs to comprehend and transform the semantic and structural information of KGs. This approach addresses the issues of missing facts and the inability to leverage semantic information from text in current KG-based recommendation methods.
\item We propose leveraging both local KG information and global KG information with LLMs. In addition, we design a retrieval-augmented method to enhance item representation learning.
\item Extensive experiments are conducted on four real-world datasets to validate the superiority of our method. Further analysis demonstrates the rationale behind our approach. 
\end{itemize}
\vspace{-0.2em}

\begin{figure*}[t]
\setlength{\belowcaptionskip}{-3mm} 
  \centering
  \includegraphics[width=0.83\textwidth]{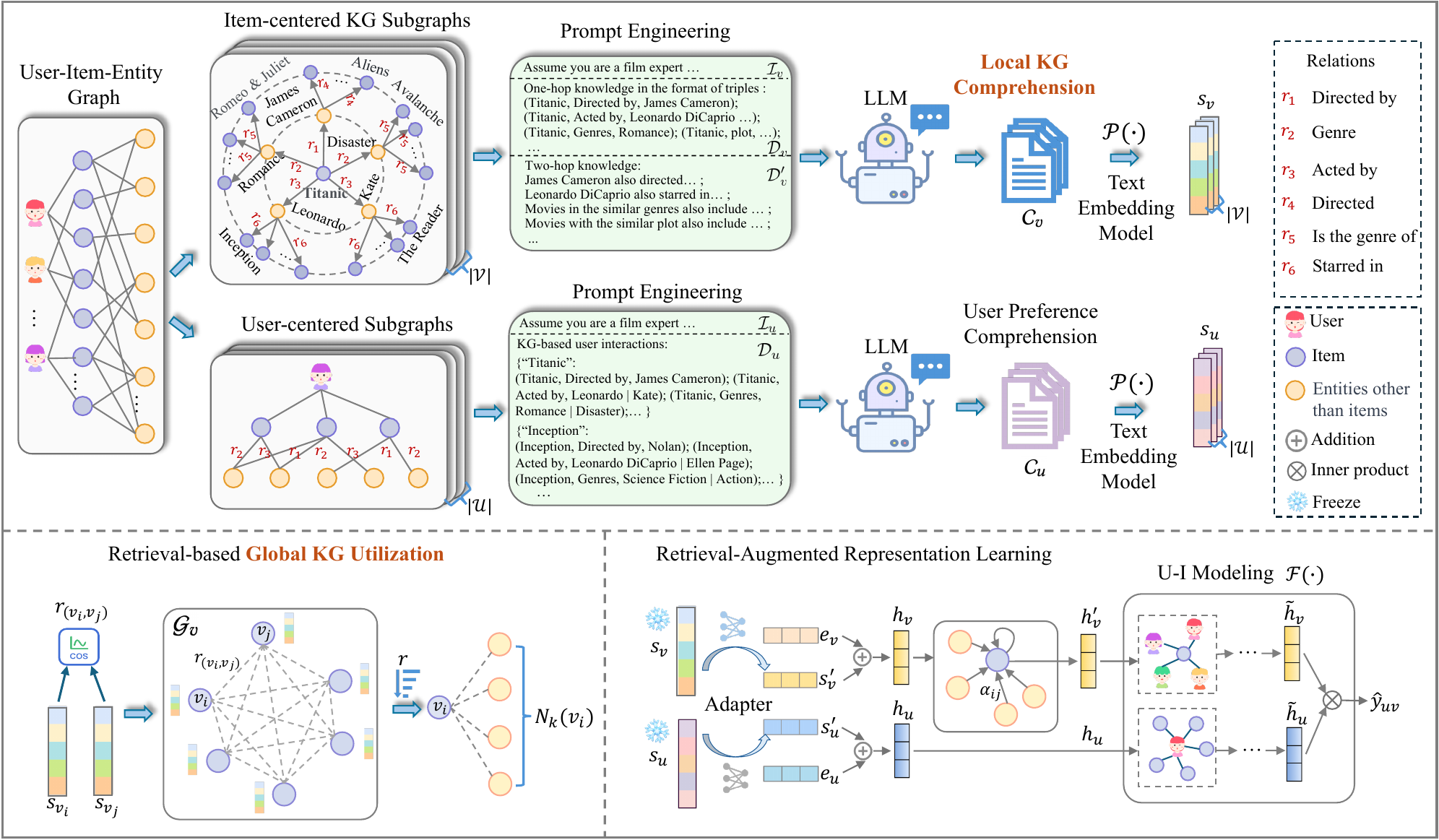}
  \caption{The framework of our proposed CoLaKG.
  } 
   \label{method_fig}
\end{figure*}
\vspace{-1em}
\section{Related Work}
\subsection{Knowledge-aware Recommendation}
Existing knowledge-aware recommendation methods can be categorized into three types \cite{guo2020survey}: embedding-based methods, path-based methods, and GNN-based methods.
Embedding-based methods \cite{cao2019unifying,zhang2016collaborative,wang2018dkn,lin2024multi} enhance the representations of users and items by leveraging the relations and entities within the KGs. Notable examples include CKE \cite{zhang2016collaborative}, which integrates various types of side information into a collaborative filtering framework using TransR \cite{lin2015learning}. Another example is DKN \cite{wang2018dkn}, which improves news representations by combining textual embeddings of sentences and knowledge-level embeddings of entities. 
Path-based methods leverage KGs to explore long-range connectivity \cite{wang2019explainable,yu2014personalized,zhang2024item,hu2018leveraging}. For example, Personalized Entity Recommendation (PER) \cite{yu2014personalized} treats a Knowledge Graph (KG) as a heterogeneous information network and extracts meta-path-based latent features to represent the connectivity between users and items along various types of relational paths. MCRec \cite{hu2018leveraging} constructs meta-paths and learns
the explicit representations of meta-paths to depict the interaction
context of user-item pairs. Despite their effectiveness, these approaches heavily rely on domain knowledge and human effort for meta-path design. Recently, GNN-based methods have been proposed, which enhance entity and relation representations by aggregating embeddings from multi-hop neighbors \cite{wang2019knowledge,wang2019kgat,wang2021learning}. For instance, KGAT \cite{wang2019kgat} employs the graph attention mechanism to propagate embeddings and utilizes multi-layer perceptrons to generate final recommendation scores in an end-to-end manner. Similarly, KGIN \cite{wang2021learning} adopts an adaptive aggregation method to capture fine-grained user intentions. Additionally, some methods \cite{yang2022knowledge,yang2023knowledge,wang2024unleashing,zhu2023knowledge} employ contrastive learning to mitigate potential knowledge noise and identify informative knowledge connections.

\vspace{-0.95em}
\subsection{LLMs for Recommendation}
In light of the emergence of large language models and their remarkable achievements in the field of NLP, scholars have begun to explore the potential application of LLMs in recommender systems \cite{wu2023survey,zhao2023survey,fan2023recommender,chen2023large}. 
Due to the powerful reasoning capabilities and extensive world knowledge of LLMs, they have been already naturally applied to zero-shot \cite{hou2024large,he2023large,wang2023zero} and few-shot recommendation scenarios \cite{bao2023tallrec,lin2024rella}. In these studies, LLMs are directly used as a recommendation model \cite{zhao2023recommender,li2023large}, where the output of LLMs is expected to offer a reasonable recommendation result \cite{wu2024survey}. However, when the dataset is sufficiently large, their performance often falls short of that achieved by traditional recommendation models. Another line of research involves leveraging LLMs as feature extractors. These methods \cite{ren2024representation,zhu2024collaborative,hu2024enhancing,wu2024exploring,acharya2023llm,kim2024large,wei2024llmrec,ren2024representation} generate intermediate decision results or semantic embeddings of users and items, which are then input into traditional recommendation models to produce the final recommendations. Unlike existing methods, our approach aims to leverage the extensive knowledge and reasoning capabilities of LLMs to understand KGs and transform them into semantic embeddings, thereby addressing existing issues in KG-based recommender systems.

\vspace{-1.0em}
\section{Preliminaries}
\begin{list}{}{\leftmargin=0em \itemindent=0em \topsep=0pt \parsep=0pt \itemsep=2pt}
\item \textbf{User-Item Interaction Graph.} 
Let \( \mathcal{U} \) and \( \mathcal{V} \) denote the user set and item set, respectively, in a recommender system. We construct a user-item bipartite graph \( \mathcal{G} = \{ (u, y_{uv}, v) | u \in \mathcal{U}, v \in \mathcal{V} \} \) to represent the collaborative signals between users and items. Here, \( y_{uv} = 1 \) if user \( u \) interacted with item \( v \), and vice versa.
\item \textbf{Knowledge Graph.}
We capture real-world knowledge about items using a heterogeneous graph composed of triplets, represented as $\mathcal{G}_k = \{(h, r, t)\}$. In this context, $h$ and $t$ are knowledge entities belonging to the set $\mathcal{E}$, while $r$ is a relation from the set $\mathcal{R}$ that links them, as exemplified by the triplet (James Cameron, directed, Titanic). Notably, the item set is a subset of the entity set, denoted as $\mathcal{V} \subset \mathcal{E}$. This form of knowledge graph enables us to model the intricate relationships between items and entities.
\item \textbf{Task Formulation.}
Following the task format of most KG-aware recommendation models, we formulate the task as follows: Given the user-item interaction graph $\mathcal{G}$ and the corresponding knowledge graph $\mathcal{G}_k$, our objective is to learn a recommendation model that predicts the probability of user $u$ interacting with item $v$.
\end{list}

\vspace{-0.5em}
\section{Methodology}
In this section, we introduce our proposed method CoLaKG in detail.
An overview of our method is illustrated in Figure \ref{method_fig}. 
CoLaKG extracts useful information from the KG at both local and global levels. By employing item-centered subgraph extraction and prompt engineering, it accurately captures the local KG. Subsequently, through retrieval-based neighbor enhancement, it supplements the current item by capturing related items from the entire KG, thereby effectively utilizing global information. The local and global information extracted by the LLM are effectively integrated into the recommendation model through a representation fusion module and a retrieval-augmented representation learning module, respectively, thereby improving recommendation performance.
\vspace{-0.3em}
\subsection{KG Comprehension with LLMs}\label{kg compreh}
Knowledge graphs have been widely utilized in recommender systems to provide semantic information and model latent associations between items. However, KGs are predominantly manually curated, leading to missing facts and limited knowledge scopes. Additionally, the highly structured nature of KGs poses challenges for utilizing textual information.
To address these limitations, we propose the use of LLMs to enhance the understanding and refinement of KGs for improved recommendations. To fully leverage KGs, our approach comprises two components: 1) Local KG Comprehension, which involves extracting item-centered KG subgraphs for analysis by LLMs, and 2) Retrieval-based Global KG Utilization, which utilizes the semantic similarity of KG subgraphs to retrieve semantically related items from the entire KG, thereby leveraging global KG information.
\begin{figure}[t]
\setlength{\abovecaptionskip}{-0.1mm} 
\setlength{\belowcaptionskip}{-5.3mm} 
  \centering  
  \includegraphics[width=0.95\columnwidth]{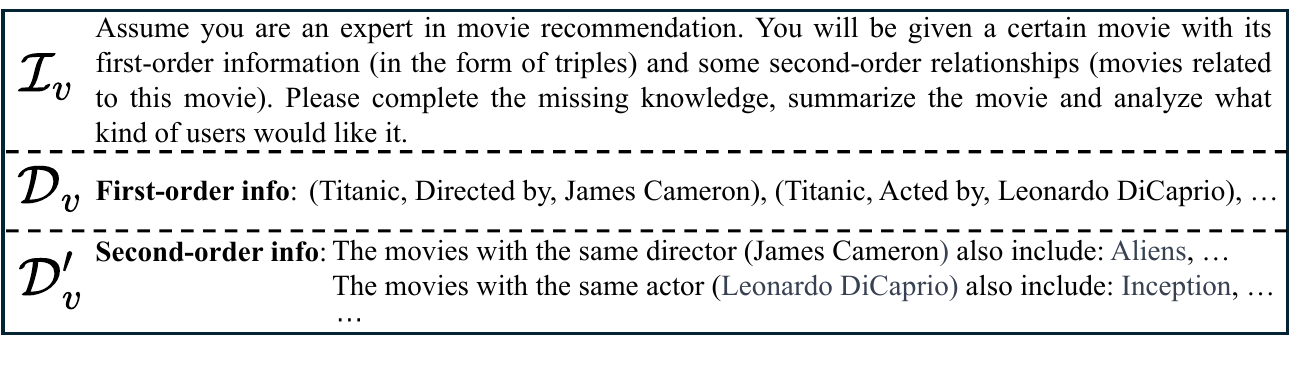}
  \caption{The prompt template for Local KG comprehension.} 
   \label{prompt_fig}
\end{figure}
\vspace{-0.3em}
\subsubsection{\textbf{Local KG Comprehension}}
\label{kg subgraph}
In this part, we introduce using LLMs to comprehend local KG information. Here, the local KG information of an item refers to the KG subgraph centered on that item, including connections within two hops.
First, we represent the first-order KG subgraph centered on each item (i.e., ego network \cite{qiu2018deepinf}) using triples. 
Specifically, given an item $v \in \mathcal{V}$, we use $\mathcal{T}_{v} =\{ {(v,r,e)|(v,r,e)} \in \mathcal{G}_k \}$ to denote the set of triplets where $v$ is the head entity. In the context of recommendation, the first-order neighboring entities of an item in a KG are usually attributes. Therefore, we use \( e \) to represent these attribute entities to distinguish them from those item entities \( v \).
During generating triples, in cases where the attribute or relation is absent, the term ``missing'' is employed as a placeholder.
Next, we consider the second-order relations in KGs. The number of entities in an ego network centered on a single entity increases exponentially with the growth of the radius. However, the input length of an LLM is strictly limited. Consequently, including all second-order neighbors associated with the central item in the prompt becomes impractical. To address this issue, we adopt a simple but effective strategy, random sampling, to explore second-order connections of \( v \). 
Let $\mathcal{E}_{v} = \{ e \mid (v, r, e) \in \mathcal{T}_{v} \}$ denote the set of first-order connected neighbors of $v$.
For each \( e \in \mathcal{E}_{v} \), we randomly sample \( m \) triples from the set \( \mathcal{T}_{e} \) to construct the triples of second-order connections, denoted as \( \mathcal{T}_{e}^m \). Here, \( \mathcal{T}_{e} = \{(e, r, v^\prime) \mid (e, r, v^\prime) \in \mathcal{G}_k, v^\prime \neq v\} \) represents the set of triples where $e \in \mathcal{E}_{v}$ is the head entity.

After converting first-order and second-order relationships into triples, we transform these triples into textual form.
For first-order relations, we concatenate all the first-order triples in $\mathcal{T}_{v}$ to form a single text, denoted as \( \mathcal{D}_{v} \). For second-order relations, we use a template to transform the second-order triples $\mathcal{T}_{e}^m$ into coherent sentences $\mathcal{D}_{v}^\prime$, facilitating the understanding of the LLM.
In addition to $\mathcal{D}_{v}$ and $\mathcal{D}_{v}^\prime$, we have carefully designed a system prompt $\mathcal{I}_v$ as the instruction to guide the generation. By combining $\mathcal{I}_v$, $\mathcal{D}_{v}$, and $\mathcal{D}_{v}^\prime$, we obtain the prompt, which is shown in Figure \ref{prompt_fig}. 
The prompt enables the LLM to fully understand, complete, and refine the KG, thereby generating the final comprehension for the $v$-centered KG subgraph. This process can be formulated as follows:
\begin{equation}
    \mathcal{C}_{v} = \text{LLMs}(\mathcal{I}_v, \mathcal{D}_{v}, \mathcal{D}_{v}^\prime).
\end{equation}
Once we have obtained the LLM's comprehension of the KG subgraphs, we need to convert these textual answers into continuous vectors for utilization in downstream recommendation models. Here, we employ a pre-trained text embedding model $\mathcal{P}$ to transform $\mathcal{C}_{v}$ into embedding vectors $\mathbf{{s}}_{v}$, which can be formulated as:
\begin{equation}
    \mathbf{{s}}_{v} = \mathcal{P}(\mathcal{C}_{v}).
\end{equation}

\subsubsection{\textbf{Rerieval-based Global KG Utilization}} \label{ii graph} 
This section introduces the utilization of global KG information. In recommendation scenarios, items that are distant in the KG can still have close semantic associations. However, the number of KG nodes increases exponentially with the number of hops, and the input length of LLMs is limited by the number of tokens. This makes it impractical to input the entire KG into an LLM directly. To address this challenge, we propose a method called retrieval-based global KG utilization.

For each item, we have obtained the semantic embedding corresponding to its local KG. Based on this, we can directly compute the semantic relationships between any two item-centered KG subgraphs. specifically, we employ the cosine similarity as a metric to quantify the relations. Given two different items $v_i$ and $v_j$, their semantical relation $r_{(v_i, v_j)}$ in the global KG is computed as:
\begin{equation} 
\label{sim}
r_{(v_i, v_j)} = \text{sim}(\mathbf{s}_{v_i}, \mathbf{s}_{v_j}), \,\,\, v_i, v_j \in \mathcal{V} \, \text{and} \, v_i \neq v_j
,
\end{equation}
where sim denotes the cosine similarity function. 
Once we obtain the semantic associations between any two items in the entire KG, we treat the semantic similarity between the two items as the edge weight between them, allowing us to construct an item-item graph:
\begin{equation}
    \mathcal{G}_v = \{ (v_i, r_{(v_i, v_j)}, v_j) | v_i, v_j \in \mathcal{V}, v_i \neq v_j \} .
\end{equation}
From a high-level perspective, we transform high-order associations between items in the KG into direct semantic connections on the constructed item-item graph $\mathcal{G}_v$. Based on this foundation, it is essential to retrieve items that are semantically strongly related to the given item, as items with lower semantic relevance may introduce noise. Specifically, given an item \(v_i\), we rank all other items \(v_j \in \mathcal{V}\) where \(v_j \neq v_i\) in descending order based on the semantic similarity \(r_{(v_i, v_j)}\). Subsequently, we retrieve the top-\(k\) items with the highest similarity scores, forming the neighbor set of $v_i$: $\mathcal{N}_k(v_i)$,
where \(0 < k < |\mathcal{V}|\) is an adjustable hyperparameter, representing the number of retrieved neighbors. 

In this manner, we filter out items with low semantic associations to the current item in the entire KG. Traditional KG-based recommendation methods aggregate items through layer-by-layer information propagation on the KG. High-order relations require many propagation and aggregation steps to be captured. In contrast, our retrieval-based method can directly recall strongly semantically related neighbors of any order across the entire KG.
The retrieved neighbors are then leveraged to enhance item representations, which will be introduced in Section \ref{item augmentation}.
\vspace{-0.5em}
\subsection{User Preference Comprehension} \label{user compreh}
The introduction of KGs allows for the expansion of user-item bipartite graphs and enables us to understand user preferences from a knowledge-driven perspective. 
Given a user \( u \), we first extract the subgraph corresponding to user \( u \) from the user-item bipartite graph, denoted as \( \mathcal{B}^u \). For each item $v \in \mathcal{B}^u $, we extract its first-order KG subgraph and represent it as a set of triples, denoted as $\mathcal{T}_{v}$. We then concatenate all triples in $\mathcal{T}_{v}$ to form a single text, denoted as $\mathcal{D}_{v}$. The detailed approach is the same as described in Section \ref{kg subgraph}.
Subsequently, we represent user $u$ with all items the user has interacted with in the training set and the corresponding knowledge triples $\mathcal{D}_{v}$:
\begin{equation} \mathcal{D}_u = \oplus_{v \in \mathcal{G}_u} \{\text{name}_v:\mathcal{D}_{v}\} ,
\end{equation}
where $\oplus$ denotes concatenation operation, and $\text{name}_v$ denotes the text name of item $v$. Additionally, we have meticulously designed a system prompt, denoted as $\mathcal{I}_u$, to serve as an instruction for guiding the generation of user preferences. By combining $\mathcal{D}_{u}$ and $\mathcal{I}_u$, we enable the LLM to comprehend the user preference for $u$, which can be formulated as:
\begin{equation}
    \mathcal{C}_{u} = \text{LLMs}(\mathcal{I}_u, \mathcal{D}_{u}).
\end{equation}
Furthermore, we also utilize the text embedding function $\mathcal{P}$ to transform the textual answers $\mathcal{C}_{u}$ into embedding vectors $\mathbf{{s}}_{u}$, which can be expressed as:
\begin{equation}
    \mathbf{{s}}_{u} = \mathcal{P}(\mathcal{C}_{u}).
\end{equation}

\subsection{Retrieval-Augmented Representation}\label{alignment}

\subsubsection{\textbf{Cross-Modal Representation Alignment}}
In a traditional recommendation model, each item and user is associated with an ID embedding. Let $\mathbf{e}_v \in \mathbb{R}^d$ represent the ID embedding of item $v$ and $\mathbf{e}_u\in \mathbb{R}^d$ represent the ID embedding of user $u$. In addition, we also obtain the semantic embedding $\mathbf{s}_v \in \mathbb{R}^{d_s}$ w.r.t. the comprehension of $v$-centric KG subgraph, and the semantic embedding $\mathbf{s}_u\in \mathbb{R}^{d_s}$ w.r.t. the comprehension of user $u$'s preference. Since ID embeddings and semantic embeddings belong to two different modalities and typically possess different embedding dimensions, we employ a learnable adapter network to align the embedding spaces. Specifically, the adapter consists of a linear map and a non-linear activation function, formulated as:
\begin{equation}
    \mathbf{s}_v^\prime = \sigma(\mathbf{W_1} \mathbf{s}_v ); \,\,\,\,\,\,
    \mathbf{s}_u^\prime = \sigma(\mathbf{W_2} \mathbf{s}_u ),
\end{equation}
where both $\mathbf{W_1}\in \mathbb{R}^{d \times d_s}$ and $\mathbf{W_2} \in \mathbb{R}^{d \times d_s}$ are are weight matrices, $\sigma$ represents the non-linear activation function ELU \cite{clevert2015fast}.
Note that during the training process, we fix \(\mathbf{s}_v\) and \(\mathbf{s}_u\), training solely the projection parameters \(\mathbf{W_1}\) and \(\mathbf{W_2}\), and the parameters of the recommendation model. 
After mapping the representations to the same space, we need to fuse the representations of the two modalities, leveraging both the collaborative signals and the semantic information to form a complementary representation. To achieve this, we employ a straightforward mean pooling technique to fuse their embeddings:
\begin{equation}
    \mathbf{h}_v = \frac{1}{2}(\mathbf{e}_v + \mathbf{s}_v^\prime); \,\,\,\,\,\,
    \mathbf{h}_u = \frac{1}{2}(\mathbf{e}_u + \mathbf{s}_u^\prime),
\end{equation}
where $\mathbf{h}_v\in \mathbb{R}^{d}$ and $\mathbf{h}_u \in \mathbb{R}^{d}$ represent the merged embeddings of item $v$ and user $u$, respectively.

\subsubsection{\textbf{Item Representation Augmentation with Retrieved Neighbors}} \label{item augmentation}
For each item, we have retrieved its semantic-related items in Section \ref{ii graph}. To fully utilize these neighbors, we propose to aggregate their information to enhance item representations. Considering the varying contributions of different neighbors to the central item, we employ the attention mechanism.
Specifically, for item $v_i$ and its top-$k$ neighbor set $\mathcal{N}_k(v_i)$, we compute attention coefficients that indicate the importance of item $v_j \in \mathcal{N}_k(v_i)$ to item $v_i$ as follows:
\begin{equation}
w_{ij}=a(\mathbf{W}\mathbf{s}_{v_i} \Vert \mathbf{W}\mathbf{s}_{v_j}).
\end{equation}
Here, $\mathbf{W} \in \mathbb{R}^{d_a \times d}$ is a learnable weight matrix to capture higher-level features of $\mathbf{s}_{v_i}$ and $\mathbf{s}_{v_j}$, $\Vert$ is the concatenation operation, $a$ denotes the attention function: $\mathbb{R}^{d_a} \times \mathbb{R}^{d_a} \rightarrow \mathbb{R}$, where we adopt a single-layer neural network and apply the LeakyReLU activation function following \cite{velivckovic2017graph}. Note that the computation of attention weights is exclusively dependent on the semantic representation of items, as our objective is to calculate the semantic associations between items, rather than the associations present in collaborative signals. In addition, we employ the softmax function for easy comparison of coefficients across different items:
\begin{equation}
    \alpha_{ij} = \text{softmax}_j(w_{ij}).
\end{equation}
The attention scores $\alpha_{ij}$ are then utilized to compute a linear combination of the corresponding neighbor embeddings. Finally, the weighted average of neighbor embeddings and the embedding of item $v_i$ itself are combined to form the final output representation for item $v_i$:
\begin{equation}
    \mathbf{h}_{v_i}^\prime = \sigma 
    \left( \frac{1}{2}
     \left(
    \mathbf{h}_{v_i} +
    \sum\nolimits_{j\in \mathcal{N}_k(v_i)} \alpha_{ij} \mathbf{h}_{v_j}
    \right) 
    \right),
\end{equation}
where $\sigma$ denotes the non-linear activation function. 

\subsection{User-Item Modeling}
Having successfully integrated the semantic information from the KG into both user and item representations, we can use them as inputs for traditional recommendation models to generate prediction results. This process can be formulated as follows:
\begin{equation}
    \hat{y}_{uv} = \mathcal{F}(\mathbf{h}_u, \mathbf{h}_{v}^\prime),
\end{equation}
where $\hat{y}_{uv}$ is the predicted probability of user $u$ interacting with item $v$, $\mathbf{h}_u$ is the representation for user $u$, $\mathbf{h}_{v}^\prime$ is the augmented representation for item $v$, and $\mathcal{F}$ denotes the function of the recommendation model.

Specifically, we select the classic model, LightGCN \cite{he2020lightgcn}, as the architecture for our recommendation method due to its simplicity and effectiveness. The trainable parameters of original LightGCN are only the embeddings of users and items, similar to standard matrix factorization. First, we adopt the simple weighted sum aggregator to learn the user-item interaction graph, which is defined as:
\begin{equation}
        \mathbf{h}_u^{(l+1)} = \sum_{v\in \mathcal{M}_u}
    \frac{1}{\sqrt{|\mathcal{M}_u|{|\mathcal{M}_v|}}}
    \mathbf{h}_v^{(l)}; \,
    \mathbf{h}_v^{(l+1)} = \sum_{u\in \mathcal{M}_v}
    \frac{1}{\sqrt{|\mathcal{M}_v||\mathcal{M}_u|}}
    \mathbf{h}_u^{(l)}    ,
\end{equation}
where $\mathbf{h}_u^{(l)}$ and $\mathbf{h}_v^{(l)}$ represent the embeddings of user $u$ and item $v$ after $l$ layers of propagation, respectively. The initial embeddings $\mathbf{h}_u^{(0)} =\mathbf{h}_u$ and $\mathbf{h}_v^{(0)} =\mathbf{h}_v^\prime$ are obtained in Section \ref{alignment}. $\mathcal{M}_u$ denotes the set of items with which user $u$ has interacted, while $\mathcal{M}_v$ signifies the set of users who have interacted with item $v$. The symmetric normalization term is given by $1/{\sqrt{|\mathcal{M}_u||\mathcal{M}_v|}}$.
Subsequently, the embeddings acquired at each layer are combined to construct the final representation:
\begin{equation}
    \tilde{\mathbf{h}}_u = \frac{1}{L+1}\sum_{l=0}^L \mathbf{h}_u^{(l)}; \,\,\,\,\,
    \tilde{\mathbf{h}}_v = \frac{1}{L+1}\sum_{l=0}^L {\mathbf{h}_v^\prime}^{(l)},
\end{equation}
where $L$ represents the number of hidden layers.
Ultimately, the model prediction is determined by the inner product of the final user and item representations:
\begin{equation}
     \hat{y}_{uv} =  {\tilde{\mathbf{h}}_u }^\top \tilde{\mathbf{h}}_v.
\end{equation}

\subsection{Model Training} \label{model train}
Our approach can be divided into two stages. In the first stage, we employ the LLM to comprehend the KGs, generating corresponding semantic embeddings for each item and user, denoted as $\mathbf{s}_v$ and $\mathbf{s}_u$, respectively. In the second stage, these semantic embeddings are integrated into the recommendation model to enhance its performance. Only the second stage necessitates supervised training, where we adopt the widely-used Bayesian Personalized Ranking (BPR) loss:
\begin{equation}
    \mathcal{L} = \sum_{(u,v^+,v^-)\in \mathcal{O}} -\text{ln} \sigma \left(
    \hat{y}_{uv^+} - \hat{y}_{uv^-}
    \right)
    + \lambda \Vert \Theta \Vert_2^2.
\end{equation}
Here, $\mathcal{O} = \{(u,v^+,v^-) | (u,v^+) \in \mathcal{R}^+, (u,v^-) \in \mathcal{R}^- \}$ represents the training set, $\mathcal{R}^+$ denotes the observed (positive) interactions between user $u$ and item $v$, while $\mathcal{R}^-$ indicates the sampled unobserved (negative) interaction set. $\sigma(\cdot)$ is the sigmoid function. $\lambda \Vert \Theta \Vert_2^2$ is the regularization term, where $\lambda$ serves as the weight coefficient and $\Theta$ constitutes the model parameter set.

\section{Experiments}
\subsection{Experimental Settings}
\begin{table}[t] 
\renewcommand\arraystretch{1.0}
	\centering
	\caption{Dataset statistics.}  
 \scalebox{0.85}{
\begin{tabular}{lrrrr}
\toprule
Statistics & MovieLens  & Last-FM  & MIND & Funds   \\
\midrule
\# Users &6,040  &1,859  &44,603  &209,999    \\
\# Items &3,260  &2,813  &15,174  &5,701    \\
\# Interactions &998,539 &86,608   &1,285,064  &1,225,318     \\
\midrule
\multicolumn{5}{c}{Knowledge Graph} \\
\midrule
\# Entities &12,068  &9,614  &32,810  &8,111    \\
\# Relations &12  &2  &14  &12    \\
\# Triples &62,958 &118,500    &307,140  &65,697     \\
\bottomrule
\end{tabular}}
\label{dataset}
\vspace{-2em}
\end{table}
\subsubsection{Datasets}
We conducted experiments on four real-world datasets, including three public datasets (MovieLens\footnote{https://grouplens.org/datasets/movielens/}, MIND\footnote{https://msnews.github.io/}, Last-FM\footnote{https://grouplens.org/datasets/hetrec-2011/}), and one industrial dataset (Fund). 
The statistics for these datasets are presented in Table \ref{dataset}. 
These datasets cover a wide range of application scenarios. Specifically, MovieLens is a well-established benchmark that collects movie ratings provided by users. MIND is a large-scale news recommendation dataset constructed from user click logs on Microsoft News. Last-FM is a well-known music recommendation dataset that includes user listening history and artist tags. The Fund dataset is sampled from the data of a large-scale online financial platform aiming to recommend funds for users. We adopt the similar setting as numerous previous studies \cite{xie2022contrastive,he2020lightgcn}, filtering out items and users with fewer than five interaction records.
For each dataset, we randomly select 80\% of each user's historical interactions to form the training set, while the remaining 20\% constitute the test set, following \cite{he2020lightgcn}. Each observed user-item interaction is considered a positive instance, and we apply a negative sampling strategy by pairing it with one negative item that the user has not interacted with.

\begin{table*}[t]  
\renewcommand\arraystretch{1.1}
	\centering
	\caption{Performance comparison of different methods, where R denotes Recall and N denotes NDCG. The best results are bolded, and the second best results are underlined. The results show our improvement is statistically significant with a significance level of 0.01.}  
	\label{tab:methodcompare}
	\setlength\tabcolsep{2.5pt}
 \scalebox{0.9}{
	\begin{tabular}{l|cccc|cccc|cccc|cccc}
		\toprule
		\multirow{2}*{Model} 
            & \multicolumn{4}{c|}{MovieLens} &\multicolumn{4}{c|}{Last-FM} 
            &\multicolumn{4}{c|}{MIND} 
            &\multicolumn{4}{c}{Funds}
            \\ 
    \cmidrule(lr){2-5}  \cmidrule(lr){6-9}  \cmidrule(lr){10-13} \cmidrule(lr){14-17} 

& R@10 & N@10 & R@20 & N@20 
& R@10 & N@10 & R@20 & N@20 
& R@10 & N@10 & R@20 & N@20
& R@10 & N@10 & R@20 & N@20
\\

\midrule 
BPR-MF &0.1257 &0.3100 &0.2048 &0.3062
       &0.1307 &0.1352 &0.1971 &0.1685
        &0.0315	&0.0238	&0.0537	&0.0310
       &0.4514 &0.3402 &0.5806 &0.3809\\
NFM  &0.1346 &0.3558 &0.2129 &0.3379
       &0.2246 &0.2327 &0.3273 &0.2830
        &0.0495	&0.0356	&0.0802	&0.0458
       &0.4388 &0.3187 &0.5756 &0.3651  \\
LightGCN  &0.1598&0.3901 &0.2512 &0.3769 
       &0.2589 &0.2799 &0.3642 &0.3321
        &0.0624	&0.0492	&0.0998	&0.0609
       &0.4992 &0.3778 &0.6353 &0.4204\\
\hline
CKE   &0.1524 &0.3783 &0.2373 &0.3609
      &0.2342 &0.2545 &0.3266 &0.3001
       &0.0526 &0.0417 &0.0822 &0.0510
      &0.4926 &0.3702 &0.6294 &0.4130
      \\
RippleNet &0.1415 &0.3669 &0.2201 &0.3423
 &0.2267 &0.2341 &0.3248 &0.2861
  &0.0472	&0.0364	&0.0785	&0.0451
&0.4764 &0.3591 &0.6124 &0.4003
\\
KGAT  &0.1536 &0.3782 &0.2451 &0.3661 
&0.2470 &0.2595 &0.3433 &0.3075
&0.0594 &0.0456 &0.0955 &0.0571
&0.5037 &0.3751 &0.6418 &0.4182
\\
KGIN  &0.1631 &0.3959 &0.2562 &0.3831
      &0.2562 &0.2742 & 0.3611 &0.3215
        &0.0640 &0.0518 &0.1022 &0.0639
      &0.5079 &0.3857 &0.6428 &0.4259
     \\
KGCL   &0.1554 &0.3797 &0.2465 &0.3677
    &\underline{0.2599} &0.2763 &\underline{0.3652} &0.3284
&\underline{0.0671}&\underline{0.0543}&\underline{0.1059}&\underline{0.0670}
    &0.5071 &0.3877 &0.6355 &0.4273
        \\
KGRec &\underline{0.1640} &\underline{0.3968} &\underline{0.2571} &\underline{0.3842} 
      &0.2571 &0.2748 &0.3617 &0.3251
      &0.0627&0.0506&0.1003&0.0625
      &\underline{0.5104} &\underline{0.3913} &\underline{0.6467} &\underline{0.4304}
      \\
\hline
RLMRec 
&0.1613&0.3920&0.2524&0.3787
&0.2597&\underline{0.2812}&0.3651&\underline{0.3335}
&0.0619&0.0486&0.0990&0.0602
&0.4988&0.3784&0.6351&0.4210 \\
KAR 
&0.1582&0.3869&0.2511&0.3722
&0.2532&0.2770&0.3612&0.3324
&0.0615&0.0480&0.1002&0.0613
&0.5033&0.3812&0.6312&0.4175 \\
CLLM4Rec
&0.1563&0.3841&0.2433&0.3637
&0.2571&0.2793&0.3642&0.3268
&0.0631&0.0494&0.1012&0.0628
&0.4996&0.3791&0.6273&0.4103 \\
\hline
\textbf{CoLaKG} 
& \textbf{0.1699} & \textbf{0.4130} & \textbf{0.2642} & \textbf{0.3974} 
& \textbf{0.2738} & \textbf{0.2948} & \textbf{0.3803} & \textbf{0.3471} 
& \textbf{0.0698} & \textbf{0.0562} & \textbf{0.1087} & \textbf{0.0684} 
& \textbf{0.5273} & \textbf{0.4012} & \textbf{0.6524} & \textbf{0.4392} 
\\
\bottomrule
\end{tabular} }
\label{comparison results}
\end{table*}

\begin{table}[h]
\centering
\caption{Validation of the generalizability of our method: Experimental results of integrating CoLaKG with various recommendation backbones.}
\scalebox{0.9}{
\begin{tabular}{l@{\hskip 0.1in}*{8}{>{\centering\arraybackslash}p{0.7cm}}}
\toprule
\multirow{2}{*}{Model} & \multicolumn{2}{c}{MovieLens} & \multicolumn{2}{c}{Last-FM} & \multicolumn{2}{c}{MIND}  \\
\cmidrule(r){2-3} \cmidrule(r){4-5} \cmidrule(r){6-7}
 & R@20 & N@20 & R@20 & N@20 & R@20 & N@20  \\
\midrule
BPR-MF &0.2048 & 0.3062 & 0.1971 & 0.1685 & 0.0537 & 0.0310 
 \\
BPR-MF+Ours & 0.2213 & 0.3255 & 0.2104 & 0.1812	 & 0.0609	 & 0.3986
 \\
\cmidrule(r){1-7}
NFM & 0.2129 & 0.3379 & 0.3273 & 0.2830 & 0.0802	 & 0.0458  \\
NFM+Ours & 0.2285 &0.3527 & 0.3478 & 0.2996 &0.0859  & 0.0487  \\
\cmidrule(r){1-7}
LightGCN & 0.2512 & 0.3769 & 0.3642 & 0.3321 & 0.0998 & 0.0609  \\
LightGCN+Ours & 0.2642 & 0.3974 & 0.3803 & 0.3471 & 0.1087 & 0.0684  \\
\bottomrule
\end{tabular}

\label{backbones}
}
\end{table}

\subsubsection{Evaluation Metrics}
To evaluate the performance of the models, we employ widely recognized evaluation metrics: Recall and Normalized Discounted Cumulative Gain (NDCG), and report values of Recall@k and NDCG@k for k=10 and 20, following \cite{he2020lightgcn,wang2019kgat}. To ensure unbiased evaluation, we adopt the all-ranking protocol. All items that are not interacted by a user are the candidates.

\subsubsection{Baseline Methods}
To ensure a comprehensive assessment,
we compare our method with 12 baseline methods, which can be divided into three categories: classical methods (BPR-MF, NFM, LightGCN), KG-enhanced methods (CKE, RippleNet, KGAT, KGIN, KGCL, KGRec), and LLM-based methods (RLMRec, KAR, CLLM4Rec). 
\begin{list}{}{\leftmargin=0em \itemindent=0em \topsep=0pt \parsep=0pt \itemsep=2pt}
\item \textbf{BPR-MF} \cite{rendle2012bpr} employs matrix factorization to model users
and items, and uses the pairwise Bayesian Personalized Ranking (BPR) loss to optimize the model.
\item \textbf{NFM} \cite{he2017neural} is an advanced factorization model that subsumes FM \cite{rendle2011fast} under neural networks.
\item \textbf{LightGCN} \cite{he2020lightgcn} facilitates message propagation between users and items by simplifying GCN \cite{kipf2016semi}.
\item \textbf{CKE} \cite{zhang2016collaborative} is an embedding-based method that uses TransR to guide entity representation in KGs to enhance performance.
\item \textbf{RippleNet} \cite{wang2018ripplenet} automatically discovers users’ hierarchical interests by iteratively propagating users’ preferences in the KG.
\item \textbf{KGAT} \cite{wang2019kgat} designs an attentive message passing scheme over the knowledge-aware collaborative graph for node embedding fusion.
\item \textbf{KGIN} \cite{wang2021learning} adopts an adaptive aggregation method to capture fine-grained user intentions.
\item \textbf{KGCL} \cite{yang2022knowledge} uses contrastive learning for knowledge graphs to reduce potential noise and guide user preference learning.
\item \textbf{KGRec} \cite{yang2023knowledge} is a state-of-the-art KG-based recommendation model which devises a self-supervised rationalization method to identify informative knowledge connections.
\item \textbf{RLMRec} \cite{ren2024representation} is an LLM-based model. It directly utilizes LLMs to generate text profiles and combine them with recommendation models through contrastive learning. Since their method is model-agnostic, to ensure a fair comparison, we chose LightGCN as its backbone model, consistent with our method. 
\item \textbf{KAR} \cite{xi2024towards} is an LLM-based model, which utilizes LLMs to enhance recommender systems with open-world knowledge
\item \textbf{CLLM4Rec} \cite{zhu2024collaborative} is also an LLM-based method. It combines ID information and semantic information and uses the LLM directly as the recommender to generate the recommendation results.
\end{list}

\subsubsection{Implementation Details} 
We implement all baseline methods according to their released code. The embedding size $d$ for all recommendation methods is set to 64 for a fair comparison. All experiments are conducted with a single V100 GPU.
We set the batch size to 1024 for the Last-FM dataset and 4096 for the other datasets to expedite training. The Dropout rate is chosen from the set \{0.2, 0.4, 0.6, 0.8\} for both the embedding layer and the hidden layers. We employ the Adam optimizer with a learning rate of 0.001. 
The maximum number of epochs is set to 2000. The number of hidden layers for the recommendation model $L$ is set to 3.
For the LLM, we select DeepSeek-V2, a robust large language model that demonstrates exceptional performance on both standard benchmarks and open-ended generation evaluations. For more detailed information about DeepSeek, please refer to their official website\footnote{https://github.com/deepseek-ai/DeepSeek-V2}. Specifically, we utilize DeepSeek-V2 by invoking its API\footnote{https://api-docs.deepseek.com/}. To reduce text randomness of the LLM, we set the temperature $\tau$ to 0 and the top-$p$ to 0.001. 
For the text embedding model $\mathcal{P}$, we use the pre-trained sup-simcse-roberta-large\footnote{https://huggingface.co/princeton-nlp/sup-simcse-roberta-large} \cite{gao2021simcse}. We use identical settings for the baselines that also involve LLMs and text embeddings to ensure fairness in comparison.



\subsection{Comparison Results}

We compare 12 baseline methods across four datasets and run each experiment five times. The average results are reported in Table \ref{comparison results}. Based on the results, we make the following observations:
\begin{itemize}[leftmargin=1em,topsep=1pt,parsep=1pt]
\item Our method consistently outperforms all the baseline models across all four datasets. The performance ceiling of traditional methods (BPR-MF, NFM, LightGCN) is generally lower than that of KG-based methods, as the former rely solely on collaborative signals without incorporating semantic knowledge. However, some KG-based methods do not perform as well as LightGCN, indicating that effectively leveraging KG is a challenging task.
\item Among the KG-based baselines, KGCL and KGRec are notable for incorporating self-supervised learning into general KG-based recommendation frameworks. However, they struggle with missing facts, understanding semantic information, and modeling higher-order item associations within the KG. In contrast, our method leverages LLMs to address these challenges without requiring self-supervised tasks, leading to significant improvements across all datasets and metrics.
\item For LLM-based recommendation methods, we have selected several representative approaches: RLMRec, KAR, and CLLM4Rec. It is evident that these methods exhibit only marginal improvements over traditional techniques. In contrast, our method demonstrates a substantial performance enhancement compared to these LLM-based baselines, thereby further validating the superiority of our approach. Our method effectively leverages LLMs to comprehend both the local subgraphs and global relationships within knowledge graphs, resulting in significant performance improvements.

\end{itemize}

\vspace{-0.5em}
\subsection{Validation of the Generalizability}
In this section, we validate the versatility of CoLaKG. Specifically, we integrate our method into three different classical recommendation model backbones and observe the performance improvements of these models across three public datasets. The results, as shown in Table \ref{backbones}, clearly indicate significant performance improvements when our method is combined with various recommendation backbones. This experiment demonstrates that our approach, which leverages LLMs to understand KGs and enhance recommendation models, can be flexibly applied to different recommendation models to improve their performance.

\begin{table}[t]
\renewcommand\arraystretch{1.0}
\centering
\caption{Ablation study on all four datasets.}
\setlength\tabcolsep{3.5pt}
\scalebox{0.9}{
\begin{tabular}{l|c|ccccc}
\toprule
 & Metric & {\makecell[c]{w/o $\mathbf{s}_v$}} & {\makecell[c]{w/o $\mathbf{s}_u$}}  & {\makecell[c]{w/o $\mathcal{N}_k(v)$}} &{\makecell[c]{w/o  $\mathcal{D}_{v}^\prime$}}& CoLaKG  \\
\midrule
\multirow{2}*{ML}  &R@20 &0.2553 &0.2613  &0.2603 &0.2628&\textbf{0.2642}  \\
  &N@20  &0.3811 &0.3948 &0.3902 &0.3960 &\textbf{0.3974}  \\
  \midrule
\multirow{2}*{Last-FM} &R@20 &0.3628 &0.3785  &0.3725 &0.3789 &\textbf{0.3803}   \\
  &N@20 &0.3278 &0.3465  &0.3403 &0.3459 &\textbf{0.3471}  \\
  \midrule
\multirow{2}*{MIND} &R@20 &0.1043 &0.1048 &0.1064&0.1076  &\textbf{0.1087}  \\
  &N@20 &0.0640 &0.0658 &0.0662&0.0671  &\textbf{0.0684}  \\
  \midrule
\multirow{2}*{Funds} &R@20 &0.6382 &0.6481 &0.6455 &0.6499 &\textbf{0.6524}  \\
  &N@20 &0.4247 &0.4351 &0.4305 &0.4378 &\textbf{0.4392}   \\

\bottomrule
\end{tabular}}
\label{ablation study}
\vspace{-0.5em}
\end{table}

\vspace{-0.5em}
\subsection{Ablation Study}
In this section, we demonstrate the effectiveness of our model by comparing its performance with four different versions across all four datasets. The results are shown in Table \ref{ablation study}, where ``w/o $\mathbf{s}_v$'' denotes removing the semantic embeddings of items, ``w/o $\mathbf{s}_u$'' denotes removing the semantic embeddings of users, ``w/o $\mathcal{N}_k(v)$'' means removing the neighbor augmentation of items based on the constructed item-item graph, and ``w/o $\mathcal{D}_{v}^\prime$'' means removing the second-order triples from the LLM's prompts.
When the semantic embeddings of items are removed, the model's performance significantly decreases across all datasets, underscoring the critical role of semantic information captured by LLMs from the KG. Similarly, the removal of user semantic embeddings also results in a performance decline, affirming that LLMs can effectively infer user preferences from the KG.
Furthermore, removing $\mathcal{N}_k(v)$ leads to a performance drop across all datasets, highlighting the significance of the item representation augmentation module based on the constructed semantic-relational item-item graph. Without this module, the model can only capture local KG information from item-centered subgraphs and cannot leverage the semantic relations present in the global KG. The inclusion of this module facilitates the effective integration of both local and global KG information.
Lastly, removing second-order KG triples from the prompts causes a slight performance decline. This finding suggests that incorporating second-order information from the KG allows the LLMs to produce a higher-quality comprehension of the local KG.

\begin{figure}[t]
\setlength{\abovecaptionskip}{-0.1mm} 
\setlength{\belowcaptionskip}{-5mm} 
  \centering
  \includegraphics[width=0.8\columnwidth]{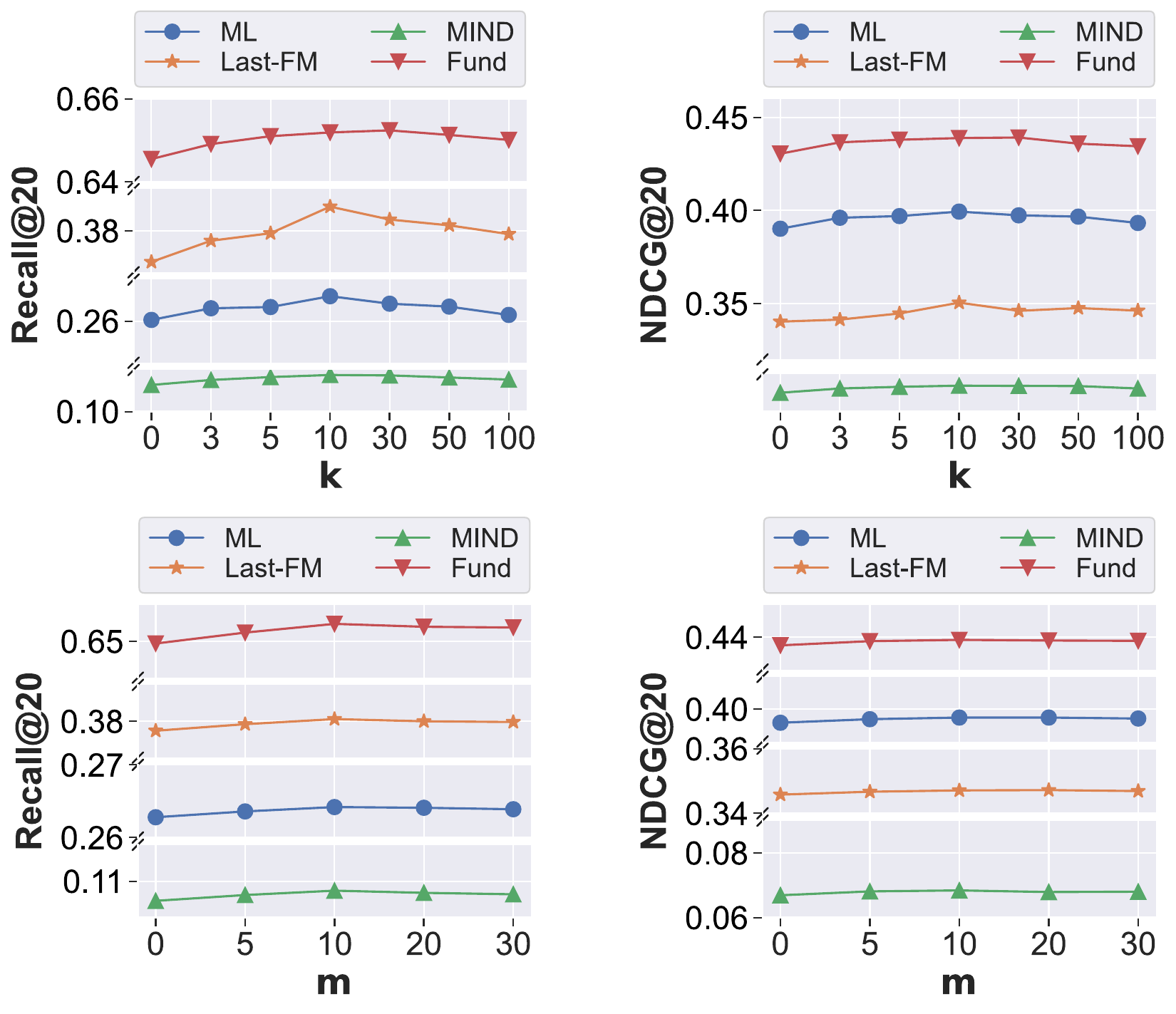}
  \caption{Hyperparameter study of the number of
retrieved neighbors ($k$) and sampled number of 2-hop items within the prompt ($m$) on four datasets.} 
   \label{param_study}
\end{figure}

\vspace{-0.5em}
\subsection{Hyperparameter Study}

In this section, we investigate the impact of the hyperparameter $k$ and $m$ across four datasets. Here, $k$ represents the number of semantically related neighbors, as defined in Section \ref{ii graph}, $m$ is the number of second-order neighbors used in the prompt, which is defined in Section \ref{kg subgraph}. The results are presented in Figure \ref{param_study}.
We observe that as $k$ increases, both Recall@20 and NDCG@20 initially rise and then slightly decline across all datasets. The performance is worst when $k=0$ and best when $k$ is between 10 and 30. When $k=0$, no neighbors are used, which is equivalent to the ablation study without $\mathcal{N}_k(v)$, thereby not incorporating any global semantic associations from the KG. When $k>0$, the introduction of semantically related items enhances the item's representations, leading to a noticeable improvement. However, as $k$ continues to increase, some noise may be introduced because the relevance of neighbors decreases with their ranking. Consequently, items with lower relevance may interfere with the recommendation performance. Our findings suggest that a range of 10-30 neighbors is optimal. As the value of \( m \) increases, the metrics initially rise and then slightly decline. When \( m = 0 \), the prompt used to understand the KG subgraph includes only first-order neighbors, resulting in the poorest performance. This indicates the positive impact of incorporating second-order neighbors. However, as \( m \) continues to grow, the marginal benefits diminish, and additional noise may be introduced, leading to a slight decrease in performance. 

\vspace{-0.5em}
\subsection{Robustness to Varying Degrees of Sparsity}
\begin{figure}[t]
\setlength{\abovecaptionskip}{-0.1mm} 
\setlength{\belowcaptionskip}{-5mm} 
  \centering
  \includegraphics[width=0.9\columnwidth]{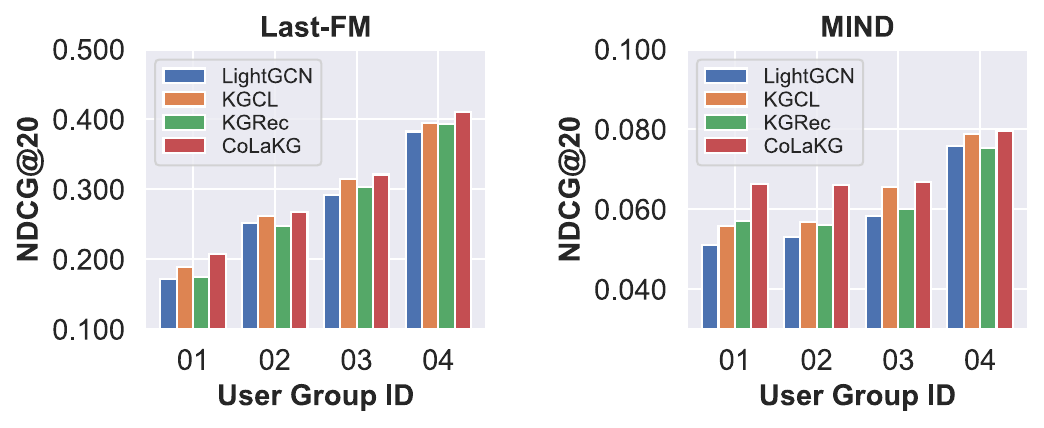}
  \caption{Performance comparison on different user groups, where a smaller group ID indicates fewer interaction records.} 
   \label{group study}
\end{figure}
One of the key functions of KGs is to alleviate the issue of data sparsity. To further examine the robustness of our model against users with varying levels of activity, particularly its performance with less active users, we sort users based on their interaction frequency and divide them into four equal groups. A lower group ID indicates lower user activity (01 being the lowest, 04 the highest). We analyze the evaluation results on two relatively sparse datasets, Last-FM and MIND, as shown in Figure \ref{group study}.
By comparing our model with three representative and strong baseline models, we observe that our model consistently outperforms the baselines in each user group. Notably, the improvement ratio of our model in the sparser groups (01 and 02) is higher compared to the denser groups (03 and 04). For the group with the most limited data (Group 01), our model achieves the most significant lead. This indicates that the average improvement of our model is primarily driven by enhancements in the sparser groups, demonstrating the positive impact of CoLaKG in addressing data sparsity.

\begin{figure}[t]
\setlength{\abovecaptionskip}{-0.1mm} 
\setlength{\belowcaptionskip}{-5mm} 
  \centering
  \includegraphics[width=0.88\columnwidth]{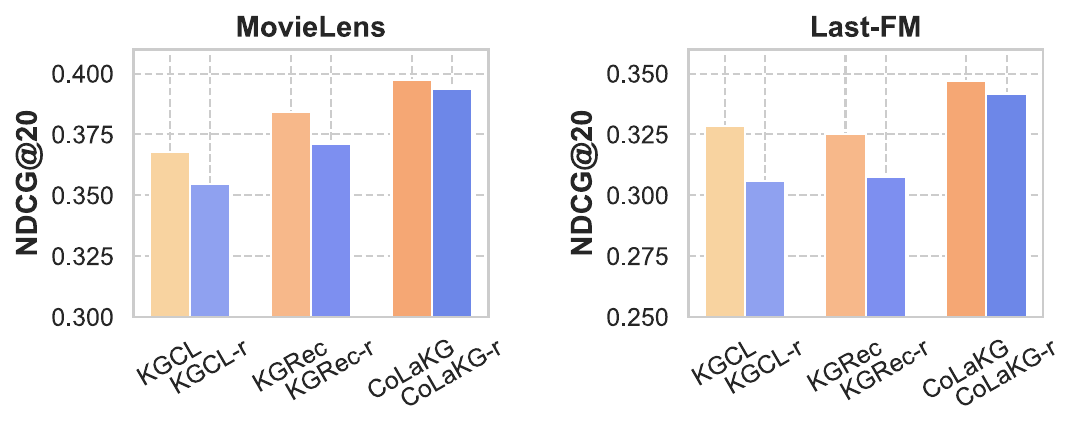}
  \caption{Performance comparison using complete KG and incomplete KG on two datasets. Models with the "-r" suffix indicate their performance in the presence of missing facts.} 
   \label{remove_facts}
\end{figure}

\vspace{-0.55em}
\subsection{Robustness to Missing Facts}
To further demonstrate the robustness of our proposed ColaKG in scenarios where facing the challenge of missing facts in KG, we conducted comparative experiments on the Movielens and Last-FM datasets. Specifically, we randomly dropped 30\% of the KG entities and relations to construct datasets with significant missing facts based on the original datasets. The comparison results of our method with representative KG-based baselines on the constructed datasets are shown in Figure \ref{remove_facts}. From the experimental results, we can see that removing KG facts has a negative impact on all KG-based methods. However, in the case of an incomplete KG, our method, CoLaKG, still outperforms the other baselines. Furthermore, by comparing the performance degradation of different methods on incomplete and complete KGs, we find that our method experiences the smallest proportion of performance decline. This demonstrates that our method can effectively mitigate the impact of missing facts on performance, further proving the robustness and superiority of the proposed approach.

\vspace{-0.55em}
\subsection{Case Study} 
\begin{figure}[t]
\setlength{\abovecaptionskip}{-0.1mm} 
\setlength{\belowcaptionskip}{-6mm} 
  \centering
  \includegraphics[width=0.95\columnwidth]{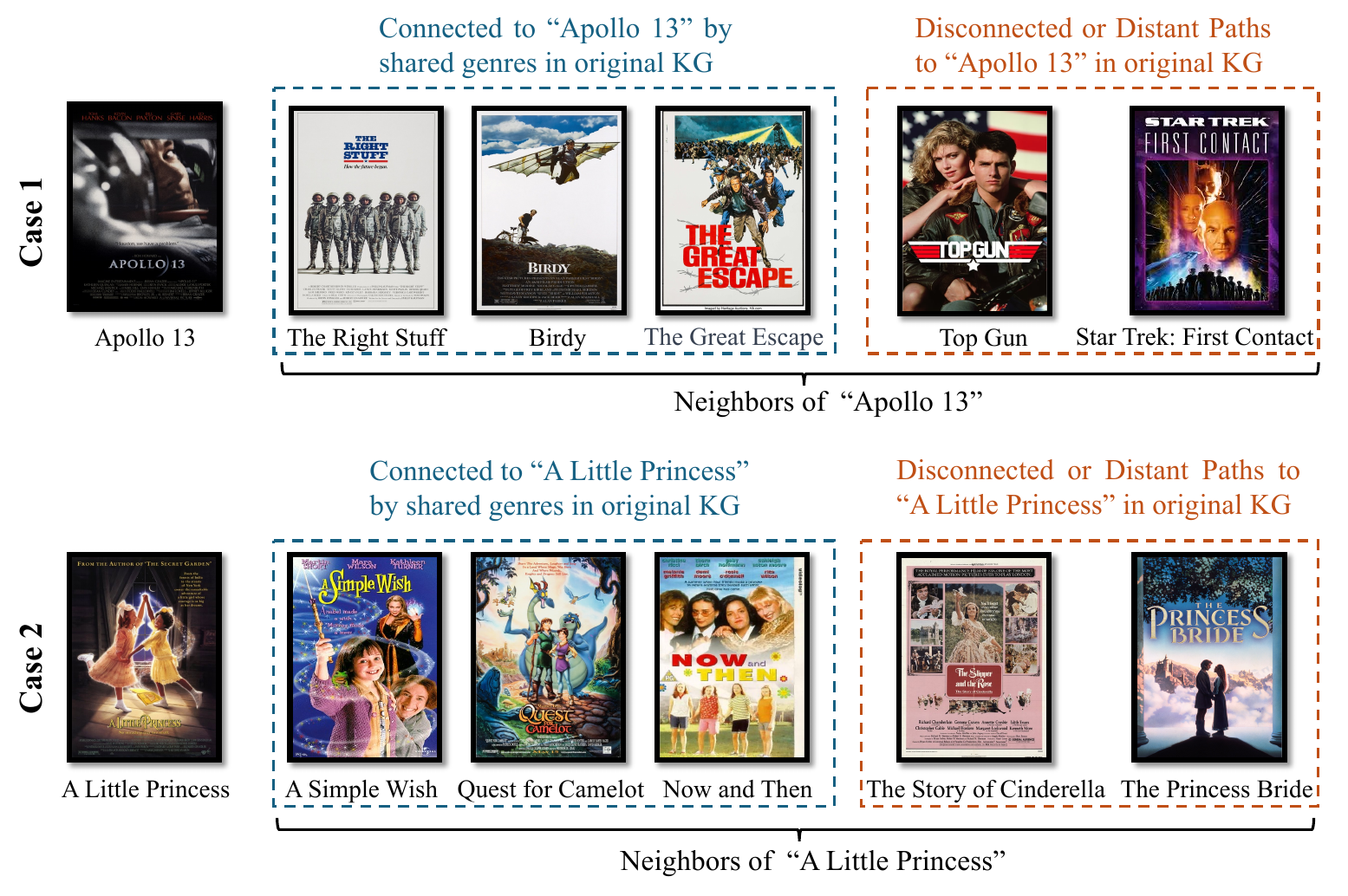}
  \caption{Case study.} 
   \label{case study}
\end{figure}

In this section, we conduct an in-depth analysis of the rationality of our method through two real cases. 
In the first case, we present the movie ``Apollo 13'' and its five semantically related neighbor items in the item-item graph identified by our method. The first three movies belong to the same genre as ``Apollo 13'', making them 2-hop neighbors in the KG. In contrast, the other two movies, ``Top Gun'' and ``Star Trek'', do not share any genre or other attributes with ``Apollo 13'', indicating they are distant or unconnected in the KG. However, ``Top Gun'' and ``Star Trek'' are semantically related to ``Apollo 13'' as they all highlight themes of human resilience, courage, and the spirit of adventure. Traditional KG-based recommendation methods, which rely on layer-by-layer information propagation, struggle to capture such high-order neighbors. In contrast, our method leverages similarity calculations based on item-centered KG semantic embeddings, successfully identifying these two strongly related movies. This demonstrates that our approach can effectively and efficiently capture semantically relevant information from the global KG.
In the second case, we examine the movie ``A Little Princess'' and its related neighbors. Among the five related movies identified, ``The Story of Cinderella'' and ``The Princess Bride'' should share the same genre as ``A Little Princess''. However, due to missing genres in the KG, these movies lack a path to ``A Little Princess'' within the KG. Despite this, our method successfully identifies these two movies. This demonstrates that our approach, by leveraging LLMs to complete and interpret the KG, can effectively address challenges posed by missing key attributes.

\vspace{-0.55em}
\section{Conclusion}
In this paper, we analyze the limitations of existing KG-based recommendation methods and propose a novel approach, CoLaKG, to address these issues. CoLaKG comprehends item-centered KG subgraphs to obtain semantic embeddings for both items and users. These semantic embeddings are then used to construct a semantic relational item-item graph, effectively leveraging global KG information. We conducted extensive experiments on four datasets to validate the effectiveness and robustness of our method. The results demonstrate that our approach significantly enhances the performance of recommendation models.

\vspace{-0.55em}
\section{Acknowledgments}
This work was supported by the Early Career Scheme (No. CityU 21219323) and the General Research Fund (No. CityU 11220324) of the University Grants Committee (UGC), and the NSFC Young Scientists Fund (No. 9240127).

\bibliographystyle{ACM-Reference-Format}
\bibliography{reference}

\end{document}